\documentclass[12pt]{article}
\usepackage{amsmath, amssymb, mathrsfs, bbm}
\usepackage{graphicx}
\usepackage{dcolumn}
\usepackage{bm}
\usepackage{amsfonts}
\usepackage{graphicx}
\usepackage{comment}
\usepackage{hyperref}

\usepackage{bbm}

\newcommand{\beqs}{\begin{equation*}}
\def\beq{\begin{equation}}
\newcommand{\eeqs}{\end{equation*}}
\def\eeq{\end{equation}}
\newcommand{\beqas}{\begin{eqnarray*}}
\newcommand{\beqa}{\begin{eqnarray}}
\newcommand{\eeqas}{\end{eqnarray*}}
\newcommand{\eeqa}{\end{eqnarray}}
\newcommand{\dps}{\displaystyle}
\newcommand{\eps}{\varepsilon}
\newcommand{\al}{\alpha}

\newcommand{\de}{\delta}

\newcommand{\la}{\lambda}
\newcommand{\si}{\sigma}

\newcommand{\Om}{\Omega}

\newcommand{\La}{\Lambda}

\def\cF{{\cal F}}

\newcommand{\bea}{\begin{eqnarray}}
\newcommand{\eea}{\end{eqnarray}}
\newcommand{\nn}{\nonumber}

\def \del {\partial}
\def \dels {\partial\kern-.5em / \kern.5em}

\def \eps {\epsilon}

\def \lam {\lambda}

\def \Om {\Omega}

\def \rr {\raise.35ex\hbox{\small $\prime$}\kern-.17em{\mbox{\large $\imath$}}}
\def \II {I\hspace{-.1em}I\hspace{.1em}}

\def \IIB {\mbox{\II B\hspace{.2em}}}
\def \As {{A\kern-.5em / \kern.5em}}
\def \Ds {D\kern-.7em / \kern.5em}

\begin{document}

\begin{titlepage}

\begin{center}
\textbf{\LARGE Lagrangian Formulations of\\ 
Self-dual Gauge Theories in 
\\Diverse Dimensions\\
}

{\large
\vskip .5in
Wei-Ming Chen
\footnote{tainist@gmail.com}, 
Pei-Ming Ho
\footnote{pmho@phys.ntu.edu.tw}
}
\vskip 3mm
{\it\large
Department of Physics, 
Center for Theoretical Sciences \\
and Leung Center for Cosmology and Particle Astrophysics, \\
National Taiwan University, Taipei 10617, Taiwan,
R.O.C.\\}

\vspace{60pt}

~\\
\textbf{Abstract}

\end{center}

\par In this work, we study 
Lagrangian formulations for self-dual gauge theories, 
also known as chiral $n$-form gauge theories, 
for $n = 2p$ in $D = 4p+2$ dimensional spacetime.
Motivated by a recent formulation of M5-branes derived from the BLG model, 
we generalize the earlier Lagrangian formulation based on a decomposition of spacetime 
into $(D-1)$ dimensions plus a special dimension, 
to construct Lagrangian formulations based on 
a generic decomposition of spacetime into $D'$ and $D'' = D - D'$ dimensions.
Although the Lorentz symmetry is not manifest, 
we prove that the action is invariant under modified Lorentz transformations.

\end{titlepage}

\setcounter{footnote}{0}

\section{Introduction}

Self-dual gauge theories, or chiral theories, are full of both physical and mathematical interests. 
The goal of this work is to provide new Lagrangian formulations of 
self-dual gauge theories for 
spacetime dimensions equal to 2 modulo 4, i.e., $D = 4p+2$, 
for $p = 0, 1, 2, 3, \cdots$.
This includes chiral bosons in 2D, 
self-dual 3-form gauge field for M5-brane in M theory 
and self-dual 5-form gauge field in type \IIB superstring theory. 
For simplicity we will assume that the spacetime is Minkowski space. 
It should be straightforward to generalize the formulation to curved spacetime
with Lorentzian signature. 

Let us recall that the dual of a tensor $T$ in $D$ dimensions is defined as
\beq
\tilde{T}_{\mu_1\cdots\mu_k} \equiv 
\frac{1}{(D-k)!} \epsilon_{\mu_1\cdots\mu_D} T^{\mu_{k+1}\cdots\mu_D}. 
\eeq
The self-duality condition of a gauge field is
\beq
{\cal F} \equiv F - \tilde{F} = 0.
\eeq
Apparently the field strength must be a $D/2$-form, 
and $D$ must be even. 
Self-duality conditions are not consistent 
in $D = 4p$ ($p = 1, 2, 3, \cdots$) dimensional Minkowski space,
and they will not be considered in this work. 

It is well known that, without auxiliary fields, 
a manifestly Lorentz invariant action cannot be found for self-dual gauge theory. 
In the literature on self-dual theories \cite{SD,FJ}, 
a special (but arbitrary) direction has to be singled out 
to write down a Lagrangian. 
Recently, in the study of M theory, 
a new Lagrangian formulation of M5-brane in large $C$-field background
\cite{Ho:2008nn,Ho:2008ve,Ho:2009zt},
which is a self-dual gauge field theory in 6 dimensions, 
was derived from the BLG model \cite{BL,Gustavsson} for multiple M2-branes. 
In this formulation, the 6 coordinates of the base space 
are divided into two sets of 3 coordinates 
$\{x^{a}\}_{a = 1}^3$ and $\{x^{\dot{a}}\}_{\dot{a} = 1}^3$, 
in contrast with the old formulation of M5-branes \cite{oldM5}, 
which uses a decomposition of base space coordinates into two sets of 1 and 5 coordinates. 
While both decompositions 6 = 1+5 and 6 = 3+3 admit 
Lagrangian formulations of self-dual theories, 
the natural question is: 
does there exist a formulation corresponding to the decomposition 6 = 2+4? 
More generally, for a self-dual gauge theory in $D = 4p+2$ dimensions 
(so the gauge field strength is a $(2p+1)$-form), 
can we find a Lagrangian formulation for all possible decompositions $D = D' + D''$?

The answer to the question above is yes. 
In the following,
we provide new Lagrangian formulations for arbitrary spacetime decompositions. 
The action is given in section 2, 
its gauge symmetry in section 3. 
The proof that the theory is a theory of self-dual gauge fields is given separately 
for three classes of decompositions: (i) $D' = D''$ (section 4), 
(ii) $D'' > D' = $ odd (section 5) 
and (iii) $D'' > D' = $ even (section 6). 
We show in section 7 
that although the action is no longer manifestly invariant 
under those Lorentz transformations which mix the two sets of coordinates in the decomposition, 
the action is invariant under certain modified Lorentz transformation laws.
In section 8, we give the interaction term in the action to describe 
the coupling of the gauge field to a charged $(2p-1)$-brane. 
Explicit examples for $(D', D'') = (1, 1)$, $(1, 5)$, $(2, 4)$, $(3, 3)$ are given in section 9. 
We point out the relationship between our result and 
the holographic action of Belov and Moore \cite{BelovMoore} in section 10 
and our conclusion will be given in section 11. 

\section{Action}
\label{Action}

We decompose the $D$-dimensional Minkowski spacetime as 
a product space
${\cal M}^D = {\cal M}_1^{D'} \times {\cal M}_2^{D''}$.
Correspondingly, the spacetime coordinates $\{ x^{\mu} | \mu = 1, \cdots, D \}$
are divided into two sets 
\beq
{\cal M}_1^{D'} : \{ x^a | a = 1, \cdots, D' \} \qquad \mbox{and} \qquad 
{\cal M}_2^{D''} : \{ x^{\dot a} | \dot{a} = 1, \cdots, D'' \}. 
\eeq 
We assume that $D' \leq D''$, so that $1 \leq D' \leq D/2$. 
This is just a convention except that when $D'$ is even, 
the signature of ${\cal M}_1^{D'}$ must be Lorentzian, 
and the signature of ${\cal M}_2^{D''}$ Euclidean. 
The Lorentzian signature of spacetime can be either $(+-\cdots-)$ or $(-+\cdots+)$.
The expressions below are valid for both conventions. 

Due to the decomposition of the coordinates, 
the gauge potential $A_{\mu_1 \cdots \mu_{2p}}$
is naturally decomposed into fields of different types 
depending on the number of dotted or undotted indices as
\beq
\{ A_{a_1\dotso a_{j} \dot a_1\dotso \dot a_{2p-j}} | j=0,\dotso, D' - \al, \},
\label{Acomp}
\eeq
where 
\beq
\label{aldef}
\al = \delta_{D', D/2} 
\eeq
so that $D' - \al = \mbox{min}(D', 2p)$.
$D'$ is greater than $2p$ only for the special case $(D', D'') = (D/2, D/2)$,
and the parameters $\al$ is used to keep track of 
whether the decomposition $D = D' + D''$ 
happens to be the special case $(D', D'') = (D/2, D/2)$. 
The indices $a_1\cdots a_j$ of $A$ in (\ref{Acomp}) should be skipped altogether if $j=0$, 
and $\dot{a}_1\cdots \dot{a}_{2p-j}$ skipped altogether when $j=2p$. 
All expressions below should also be interpreted in this way.

Naturally, the totally antisymmetrized tensor in $D$ dimensions 
is naturally decomposed into a product of the totally antisymmetrized tensor 
in $D'$ and $D''$ dimensions
\beq
\epsilon_{a_1\cdots a_{D'} \dot{a}_1 \cdots \dot{a}_{D''}} = 
\epsilon_{a_1 \cdots a_{D'}} \epsilon_{\dot{a}_1\cdots \dot{a}_{D''}}.
\eeq
For an arbitrary tensor $T_{\mu_1 \cdots \mu_k \nu_1 \cdots \nu_l \mu_{k+1} \cdots \mu_m \cdots}$, 
we will use the notation 
\beq
T_{[\mu_1 \cdots \mu_k | \nu_1 \cdots \nu_l | \mu_{k+1} \cdots \mu_m] \cdots} 
\equiv \frac{1}{m!} \sum_{\sigma} \mbox{sgn}(\sigma) 
T_{\mu_{\sigma_1}\cdots \mu_{\sigma_k} \nu_1 \cdots \nu_l \mu_{\sigma_{k+1}} \cdots \mu_{\sigma_m}\cdots}, 
\eeq
where the sum $\sum_{\sigma}$ is carried out for all permutations of $m$ objects 
and $\mbox{sgn}(\sigma)$ is the signature of the permutation $\sigma$.
Note that the antisymmetrized tensor is normalized by $1/m!$, 
and that the indices sandwiched by $|\cdot|$ are not antisymmetrized.
Using this notation, we can express the field strength and its dual as
\beqa
F_{a_1\dotso a_{j} \dot a_1\dotso \dot a_{D/2-j}}&=&j\partial_{[a_1}A_{a_2\dotso a_{j}]\dot a_1\dotso \dot a_{D/2-j}}
+(D/2-j)\partial_{[\dot a_{D/2-j}}A_{|a_1\dotso a_{j}|\dot a_1\dotso \dot a_{D/2-j-1}]},\\
\tilde F_{a_1\dotso a_{j} \dot a_1\dotso \dot a_{D/2-j}}&=&\frac{(-)^{(\frac{D}{2}-j)(D^\prime-j)}
\epsilon_{a_1\dotso a_{D^\prime}}\epsilon_{\dot a_1\dotso \dot a_{D-D^\prime}}
F^{a_{j+1}\dotso a_{D^\prime}\dot a_{D/2-j+1}\dotso \dot a_{D-D^\prime}}}{(D^\prime-j)!(\frac{D}{2}-D^\prime+j)!}
\eeqa
for $j = 0, 1, \cdots, D'$.

The following two identities,
\beqa
\notag &&F_{a_1\dotso a_{j} \dot a_1\dotso \dot a_{D/2-j}}\tilde F^{a_1\dotso a_{j} \dot a_1\dotso \dot a_{D/2-j}} \\
\label{DDI1}&=&-\frac{j!(\frac{D}{2}-j)!}{(D^\prime-j)!(\frac{D}{2}-D^\prime+j)!}
F_{a_1\dotso a_{D^\prime-j}\dot a_{1}\dotso \dot a_{D/2-D^\prime+j}}
\tilde F^{a_1\dotso a_{D^\prime-j}\dot a_{1}\dotso \dot a_{D/2-D^\prime+j}}, \\
\notag &&\tilde F_{a_1\dotso a_{j} \dot a_1\dotso \dot a_{D/2-j}}\tilde F^{a_1\dotso a_{j} \dot a_1\dotso \dot a_{D/2-j}} \\
\label{DDI2}&=&-\frac{j!(\frac{D}{2}-j)!}{(D^\prime-j)!(\frac{D}{2}-D^\prime+j)!}
F_{a_1\dotso a_{D^\prime-j}\dot a_1\dotso \dot a_{D/2-D^\prime+j}}
F^{a_1\dotso a_{D^\prime-j}\dot a_1\dotso \dot a_{D/2-D^\prime+j}}, 
\eeqa
will be useful in the calculation below.
Another notation we will use below is 
\beq
{\cal F}_{\mu_1\cdots \mu_{D/2}} \equiv (F - \tilde{F})_{\mu_1\cdots \mu_{D/2}}.
\eeq

The main result of this paper is then this.
For the decomposition $D=D^\prime +D^{\prime\prime}$ of spacetime dimension $D = 4p+2$, 
the action of a self-dual field theory is 
\beqa
\notag &&S_{D^{\prime}+D^{\prime\prime}} \\
\label{action1}&=&
-\frac{1}{4}
\int d^D x\left[\begin{array}{rl}&\displaystyle
\frac{F_{\mu_1\mu_2\dotso\mu_{D/2}}F^{\mu_1\mu_2\dotso\mu_{D/2}}}{(D/2)!} \\
-&\displaystyle\sum_{j=\lceil \frac{D^{\prime}}{2}\rceil}^{D^{\prime}}
\frac{\cF_{a_1\dotso a_j\dot a_1\dotso\dot a_{D/2-j}}\cF^{a_1\dotso a_j\dot a_1\dotso\dot a_{D/2-j}}}{j!(\frac{D}{2}-j)!}
\end{array}\right] \label{S1} \\
\label{action2}&=&
-\frac{1}{2}
\int d^D x\sum_{j=\lceil \frac{D^{\prime}}{2}\rceil}^{D^{\prime}}
\frac{\tilde F_{a_1\dotso a_j\dot a_1\dotso\dot a_{D/2-j}}\cF^{a_1\dotso a_j\dot a_1\dotso\dot a_{D/2-j}}}
{j!(\frac{D}{2}-j)!2^{\de_{j, D^\prime /2}}} \label{S2} \\
\label{action3}&=&
-\frac{1}{2}
\int d^D x\sum_{j=\lceil \frac{D^{\prime}}{2}\rceil}^{D^{\prime}}
\frac{F_{a_1\dotso a_{D^{\prime}-j}\dot a_1\dotso\dot a_{D/2-D^{\prime}+j}}
\cF^{a_1\dotso a_{D^{\prime}-j}\dot a_1\dotso\dot a_{D/2-D^\prime+j}}}
{(D^{\prime}-j)!(\frac{D^{\prime\prime}-D^\prime}{2}+j)!2^{\de_{j,D^\prime /2}}}, \label{S3}
\eeqa
where we used the two identities \eqref{DDI1} and \eqref{DDI2} 
to express the action in 3 different ways for the convenience of the reader. 
In the 2nd and 3rd expressions of the action,
the factor $2^{\de_{j,D^\prime/2}}$ 
in the denominator contributes only when $D'$ is even.
In the following we will show that the equations of motion derived 
from varying this action are equivalent to the condition of self-duality, 
and that this action is invariant under a hidden Lorentz symmetry 
although it is not manifestly Lorentz invariant. 

It was pointed out in \cite{Witten} that 
the quantum theory of a self-dual gauge field is not unique 
on a generic spacetime. 
Instead there is a 1-1 correspondence between the spin structure on the base space (spacetime) 
and the path integral of a self-dual gauge field. 
The action proposed above is presumably the action 
suitable for spacetime with trivial topology 
so that the spin structure is unique. 
We leave the issue on topology and 
the definition of quantum theory for generic spacetime manifold 
to future works. 

\section{Gauge Symmetry}
\label{GaugeSymmetry}

The self-duality condition 
\beq
{\cal F}_{\mu_1\cdots\mu_{D/2}} = 0
\eeq
can be decomposed into $D'+1$ different types of equations
\beq
{\cal F}_{a_1\cdots a_j \dot{a}_1\cdots\dot{a}_{D/2-j}} = 0 \qquad 
(j = 0, 1, \cdots, D'). 
\eeq
(The quantity ${\cal F}$ vanishes identically if $j > D'$ because 
${\cal F}$ is totally antisymmetrized in all indices.)
But not all of them are independent. 
In fact, only $\lceil (D'+1)/2 \rceil$, roughly half, of them are independent.
($\lceil (D'+1)/2 \rceil$ denotes the smallest integer not smaller than $(D'+1)/2$.)
Hence the action, if it is correct, should produce $\lceil (D'+1)/2 \rceil$ different types of 
independent equations of motion that are equivalent to these conditions. 
On the other hand, the number of different types of gauge potential components 
is $(D' - \al + 1)$ (see eq.(\ref{Acomp})).
Varying them in the action should give us 
$(D' - \al + 1)$ different types of equations of motion.
Except the special case $(D', D'') = (1, 1)$, 
for which 
the number of equations of motion equals 
the number of self-duality conditions,
for all other cases, 
$(D'-\al+1) > \lceil (D'+1)/2 \rceil$.
Thus, in general, 
the equations of motion cannot be all independent. 
The number of equations of motion that are trivial should be 
\beq
(D' - \al + 1) - \lceil (D'+1)/2 \rceil = \lceil D'/2 \rceil - \al.
\label{numgaugesymm}
\eeq

It turns out that
the action (\ref{S1}-\ref{S3}) has the special feature that 
many components of the gauge potential appear only in total derivative terms. 
Therefore, in addition to the ordinary gauge symmetry 
\beq
A \rightarrow A + d \Lambda, 
\label{usualgauge}
\eeq
the action is invariant under the following gauge transformations
\beqa
\label{gtf}
\begin{array}{rll}
\de A_{a_1\dotso a_k\dot a_1\dotso\dot a_{2p-k}}=&0 & 
~(k= 0, 1, \cdots, \lfloor D'/2 \rfloor - 1), \\
\de A_{a_1\dotso a_k\dot a_1\dotso\dot a_{2p-k}}
=&f_{a_1\dotso a_k\dot a_1\dotso\dot a_{2p-k}}(x^a) & 
~(k = \lfloor D'/2 \rfloor), \\
\de A_{a_1\dotso a_k\dot a_1\dotso\dot a_{2p-k}}
=&\Phi_{a_1\dotso a_k\dot a_1\dotso\dot a_{2p-k}}(x^a, x^{\dot a}) & 
~(k=\lfloor D'/2 \rfloor +1, \dotso, D' - \al),
\end{array}
\eeqa
where the $f$'s are functions independent of all $x^{\dot a}$'s, 
and the $\Phi$'s are arbitrary functions.
By choosing $\Phi$ suitably, 
$(\lceil D'/2 \rceil - \al)$ of the $(D' - \al + 1)$ different types of $A$ fields 
can be gauged away.
Thus there are only $\lceil (D'+1)/2 \rceil$ types of nontrivial equations of motion, 
as it should be (see (\ref{numgaugesymm})).
The gauge transformation parametrized by the $f$'s in (\ref{gtf}) 
will play an important role when $D = 2$.
The fact that the action is invariant under the transformations (\ref{gtf}) 
will be checked below when we compute the equations of motion.

In the following we will show case by case that,
for (1) $D' = D''$, (2) $D' < D''$ with $D'$ odd, 
and (3) $D' < D''$ with $D'$ even, 
the space of solutions to the equations of motion defined by 
the action (\ref{action1}) is the same as the space of all self-dual field configurations. 
More specifically, 
the equations of motion in our theory 
(for those gauge potential components that cannot be gauged away)
are 2nd order differential equations 
which can be integrated into 1st order differential equations
with arbitrary functions introduced as integral "constants". 
These additional functions can be identified with 
those components of the gauge potential that are gauged away, 
or as a shift of those fields that can be absorbed by a field redefinition.
After this identification, 
the 1st order differential equations are exactly identical to  
the self-duality conditions.

Strictly speaking, the derivation given below only implies that 
every solution to the equations of motion of our theory 
correspond to a self-dual configuration, 
but it does not immediately imply that 
every self-dual configuration can find its counterpart in our theory. 
To prove that the space of solutions for our theory is really 
the same as the space of self-dual configurations, 
one needs to count the number of free components in $A$
and examine the constraints (equations of motion or 
self-duality conditions). 
We have carried out this straightforward yet tedious proof
case by case but will skip it here.

\section{$D' = D''$}
\label{D'=D''}

For the special case $D^\prime=D^{\prime\prime}=\frac{D}{2}=2p+1$ 
(and so $\al = 1$), 
the gauge transformation laws \eqref{gtf} are
\beqa
\label{gtfDD}\begin{array}{rll}
\de A^{a_1\dotso a_{2p-k}\dot a_1\dotso\dot a_{k}}=&\Phi^{a_1\dotso a_{2p-k}\dot a_1\dotso\dot a_{k}}(x^a, x^{\dot a}),
&~k=0,1,\dotso,p-1,\\
\de A^{a_1\dotso a_{2p-k}\dot a_1\dotso\dot a_{k}}=&f^{a_1\dotso a_{2p-k}\dot a_1\dotso\dot a_{k}}(x^a),
&~k=p,\\
\de A^{a_1\dotso a_{2p-k}\dot a_1\dotso\dot a_{k}}=&0,
&~k=p+1,\dotso,2p, 
\end{array}
\eeqa
in addition to the usual gauge transformation (\ref{usualgauge}). Without loss of generality, here we assume the indices $a_i$ and $\dot a_i$ correspond to the $SO(D'-1,1)$ and
$SO(D'')$ subgroup of the full Lorentz group, respectively.

The action (\ref{S2}) is 
\beqa
S=
-\frac{1}{2}
\int d^D x\sum_{j=p+1}^{2p+1}\frac{\tilde F_{a_1\dotso a_j\dot a_1\dotso\dot a_{2p-j+1}}\cF^{a_1\dotso a_j\dot a_1\dotso\dot a_{2p-j+1}}}{j!(2p-j+1)!}.
\eeqa
Its variation with respect to $A^{a_1\dotso a_{2p-k}\dot a_1\dotso\dot a_{k}}$ 
for $k=0,1,\cdots, p-1$ can be shown to vanish using the Bianchi identity
\beq
\del_{\mu_1} \tilde{F}^{\mu_1\cdots\mu_{D/2}} = 0.
\eeq

The nontrivial equations of motion are
\beqa
\frac{\de S}{\de A^{a_1\dotso a_{2p-k}\dot a_1\dotso\dot a_{k}}} = 0
\qquad (k=p,\dotso,2p).
\eeqa
For $k=p$,
\beqa
\label{sfdDD1}\frac{\de S}{\de A^{a_1\dotso a_{p}\dot a_1\dotso\dot a_{p}}}=0~\Rightarrow~
\frac{\partial^{\dot b}\cF_{a_1\dotso a_{p}\dot a_1\dotso \dot a_{p}\dot b}}{p!p!}=0,
\eeqa
and the solution to \eqref{sfdDD1} is
\beqa
\label{sfdDD1ss}\cF_{a_1\dotso a_{p}\dot a_1\dotso\dot a_{p+1}}
=\frac{(-)^{p^2+1}\epsilon_{a_1\dotso a_{2p+1}}\epsilon_{\dot a_1\dotso \dot a_{2p+1}}
\partial^{\dot a_{2p+1}}\Phi^{a_{p+1}\dotso a_{2p+1}\dot a_{p+2}\dotso \dot a_{2p}}}{(p+1)!(p-1)!}
\eeqa
for some arbitrary functions $\Phi^{a_{p+1}\dotso a_{2p+1}\dot a_{p+3}\dotso a_{2p+1}}$.
An exception is that when $p = 0$, 
the totally antisymmetrized tensor $\eps_{\dot 1}$ is trivial 
and cannot be used to write down a solution.
In this case the general solution is given by 
${\cal F}_{\dot 1} = f(x^1)$ 
for an arbitrary function independent of $x^{\dot 1}$.

If $p>0$, we can use the gauge symmetry \eqref{gtfDD} for $k=p-1$ to achieve
one of the self-duality conditions
\beqa
\label{sfdDD1s}
\cF_{a_1\dotso a_{p}\dot a_1\dotso\dot a_{p+1}}=0.
\eeqa
In case $p = 0$, we can use the second gauge transformation in (\ref{gtfDD})) 
to gauge $f$ away and arrive at the same equation.

The remaining equations of motion are,
\beqa
\label{EQMDD1}
\frac{\de S}{\de A^{a_1\dotso a_{p-1}\dot a_1\dotso\dot a_{p+1}}}=0&\Rightarrow&
\frac{\partial^b F_{ba_1\dotso a_{p-1}\dot a_1\dotso \dot a_{p+1}}
+\partial^{\dot b}F_{a_1\dotso a_{p-1}\dot a_1\dotso \dot a_{p+1}\dot b}}{(p-1)!(p+1)!}=0,\\
\notag\vdots~~~~~~~~~~&&~~~~~~~~~~\vdots \\
\label{sfdDDf}\notag
\frac{\de S}{\de A^{a_1\dotso a_{2p-k}\dot a_1\dotso\dot a_{k}}}=0&\Rightarrow&
\frac{(2p-k+1)\partial^b (\cF+F)_{ba_1\dotso a_{2p-k}\dot a_1\dotso \dot a_{k}}}{2(2p-k+1)!k!}
+\frac{(k+1)\partial^{\dot b}(\cF+F)_{a_1\dotso a_{2p-k}\dot a_1\dotso \dot a_{k}\dot b}}{2(2p-k)!(k+1)!} \\
&&=\frac{\partial^b F_{ba_1\dotso a_{2p-k}\dot a_1\dotso \dot a_{k}}
+\partial^{\dot b}F_{a_1\dotso a_{2p-k}\dot a_1\dotso \dot a_{k}\dot b}}{(2p-k)!k!}=0,\\
\notag\vdots~~~~~~~~~~&&~~~~~~~~~~\vdots\\
\frac{\de S}{\de A^{\dot a_1\dotso\dot a_{2p}}}=0&\Rightarrow&
\frac{\partial^b F_{b\dot a_1\dotso \dot a_{2p}}
+\partial^{\dot b}F_{\dot a_1\dotso \dot a_{2p}\dot b}}{(2p)!}=0.
\eeqa
Using \eqref{sfdDD1s}, 
we can rewrite the first term on the left-hand side of \eqref{EQMDD1}, i.e.,
$\del^b F_{b\cdots}$, as $\del^b \tilde{F}_{b\cdots}$, 
which equals $- \del^{\dot{b}} \tilde{F}_{\dot{b}\cdots}$ due to Bianchi identity.
Together with the 2nd term, 
the left-hand side of \eqref{EQMDD1} is a total derivative of the form 
$\del^{\dot{b}} {\cal F}_{\dot{b} \cdots}$, 
and thus \eqref{EQMDD1} can be solved in the same way we solved \eqref{sfdDD1}.
That is, we use the gauge symmetry \eqref{gtfDD} to 
absorb the functions arising due to integration, 
to obtain another self-duality condition.

The same story goes on, and each equation of motion 
is found to be of the form 
\beqa
\partial^{\dot a_{k+1}}\cF_{a_1\dotso a_{2p-k}\dot a_1\dotso \dot a_{k+1}}=0,
\qquad (k=p+1,\dotso, 2p),
\label{31}
\eeqa  
after plugging in the solution of previous equations of motion. 
For $k=p+1,\dotso,2p-1$, the solution to the equation above is 
\beqa
\label{sfdDD2}
\cF_{a_1\dotso a_{2p-k}\dot a_1\dotso \dot a_{k+1}}
=\frac{(-)^{k^2+1}\epsilon_{a_1\dotso a_{2p+1}}\epsilon_{\dot a_1\dotso \dot a_{2p+1}}
\partial^{\dot a_{2p+1}}\Phi^{a_{2p-k+1}\dotso a_{2p+1}\dot a_{k+2}\dotso \dot a_{2p}}}{(k+1)!(2p-k-1)!}
\eeqa
for some arbitrary functions $\Phi$. 
We then use the gauge symmetries \eqref{gtfDD} (for $k=0,\dotso,p-2$ in \eqref{gtfDD})
to absorb the right-hand side of \eqref{sfdDD2}.

The case of $k=2p$ is special, 
because $\cF_{\dot a_1\dotso \dot a_{2p+1}}$ 
has to be a function times the totally antisymmetrized tensor $\epsilon_{\dot a_1\dotso \dot a_{2p+1}}$.
The solution to its equation of motion is thus
\beqa
\label{sfdDD3}\cF_{\dot a_1\dotso \dot a_{2p+1}}=\epsilon_{\dot a_1\dotso \dot a_{2p+1}}f(x),
\eeqa
where $f(x)$ is a function depending only on the $D^\prime$ coordinate $x^a$, 
which can be always written as $f(x)=\partial_a f^a(x)$ 
for some functions $f^a(x)$ depending only on $x^a$.
We can then absorb the right-hand side of \eqref{sfdDD3} by a field redefinition 
$A_{a_1\dotso a_{2p}}\rightarrow A_{a_1\dotso a_{2p}}+\epsilon_{a_1\dotso a_{2p+1}}f^{a_{2p+1}}$. 
It is easy to check that this field redefinition does no spoil any of the self-duality conditions 
already satisfied. 
Hence we have shown that all self-duality conditions are satisfied
\beqa
\cF_{a_1\dotso a_{2p-k}\dot a_1\dotso \dot a_{k+1}}=0 
\qquad (k=p,\dotso,2p).
\eeqa
More precisely, since we have to make various field redefinitions and gauge transformations \eqref{gtfDD}
in order to obtain the self-duality condition, 
if the solutions of the our equations of motion are denoted by $A$, 
the corresponding self-dual gauge field configuration $A_{SD}$ is given by
\beqa
A_{SD}^{a_{1}\cdots a_{k}\dot a_{1}\cdots \dot a_{2p-k}} &=& 
A^{a_{1}\cdots a_{k}\dot a_{1}\cdots \dot a_{2p-k}} 
\qquad \mbox{for} \quad k = 0, \cdots, p, 
\label{a1} \\
A_{SD}^{a_{1}\cdots a_{k}\dot a_{1}\cdots \dot a_{2p-k}} &=& 
(A+\Phi)^{a_{1}\cdots a_{k}\dot a_{1}\cdots \dot a_{2p-k}} 
\qquad \mbox{for} \quad k = p+1, \cdots, 2p-1, 
\label{a2} \\
A_{SD}^{a_{1}\cdots a_{2p}} &=& 
(A+\Phi)^{a_{1}\cdots a_{2p}} + \epsilon^{a_1\dotso a_{2p+1}}f_{a_{2p+1}}(x^a). 
\label{a3}
\eeqa



\section{$D'' > D'$ $=$~odd}
\label{oddDp}

Consider the case $D' = 2q+1 < D''$ $(0\leq q\leq p-1)$.
The action is (\ref{S2})
\beqa
S=-\frac{1}{2}\int d^D x\sum_{j=q+1}^{2q+1}\frac{\tilde F_{a_1\dotso a_j\dot a_1\dotso\dot a_{2p-j+1}}
\cF^{a_1\dotso a_j\dot a_1\dotso\dot a_{2p-j+1}}}{j!(2p-j+1)!},
\eeqa
and the gauge transformations are (\ref{gtf})
\beqa
\label{gtfodd}\begin{array}{rll}
\de A^{a_1\dotso a_{2q-k+1}\dot a_1\dotso\dot a_{2p-2q+k-1}}&=
\Phi^{a_1\dotso a_{2q-k+1}\dot a_1\dotso\dot a_{2p-2q+k-1}}(x^a, x^{\dot a}),
&k=0,1,\dotso,q, \\
\de A^{a_1\dotso a_{2q-k+1}\dot a_1\dotso\dot a_{2p-2q+k-1}}&=
f^{a_1\dotso a_{2q-k+1}\dot a_1\dotso\dot a_{2p-2q+k-1}}(x^a),
&k=q+1, \\
\de A^{a_1\dotso a_{2q-k+1}\dot a_1\dotso\dot a_{2p-2q+k-1}}&=0,
&k=q+2,\dotso,2q+1, \\
\end{array}
\eeqa
in addition to the ordinary gauge symmetry.

The nontrivial equations of motion are
\beq
\label{VRAodd}
\frac{\de S}{\de A^{a_1\dotso a_{2q-k+1}\dot a_1\dotso\dot a_{2p-2q+k-1}}} = 0
\qquad (k=q+1,\dotso,2q+1).
\eeq
Starting with $k=q+1$, we have
\beqa
\notag
\frac{\de S}{\de A^{a_1\dotso a_{q}\dot a_1\dotso\dot a_{2p-q}}}=0~\Rightarrow~
\frac{\partial^{\dot a_{2p-q+1}}\cF_{a_1\dotso a_{q}\dot a_1\dotso\dot a_{2p-q+1}}}{q!(2p-q)!}=0,
\eeqa
and it is solved by
\beq
\cF_{a_1\dotso a_{q}\dot a_1\dotso\dot a_{2p-q+1}}=\frac{(-)^{q^2+1}
\epsilon_{a_1\dotso a_{2q+1}}\epsilon_{\dot a_1\dotso \dot a_{4p-2q+1}}
\partial^{\dot a_{4p-2q+1}}\Phi^{a_{q+1}\dotso a_{2q+1}\dot a_{2p-q+2}\dotso \dot a_{4p-2q}}}{(q+1)!(2p-q-1)!}
\eeq
for some $\Phi$.
The field $\Phi$ can be eliminated by a gauge transformation (\eqref{gtfodd} with $k=q$)
\beq
A^{a_1\cdots a_{q+1}\dot{a}_1\cdots \dot{a}_{2p-q-1}} \rightarrow 
(A+\Phi)^{a_1\cdots a_{q+1}\dot{a}_1\cdots \dot{a}_{2p-q-1}},
\eeq
so that
\beq
\label{sfdodd1}
\cF_{a_1\dotso a_{q}\dot a_1\dotso\dot a_{2p-q+1}}=0.
\eeq
For the case $q=0$ ($D' = 1$), eq.\eqref{sfdodd1} is the only self-duality condition. 
This case was already studied in the literature for spacetime dimensions 
$D = 6$ \cite{oldM5} and $D = 10$ \cite{IIB9+1}.
For $q\neq 0$, we consider the equations of motion for $k>q+1$. 
They are
\beqa
\label{defodd1}
\frac{\de S}{\de A^{a_1\dotso a_{q-1}\dot a_1\dotso\dot a_{2p-q+1}}}=0
&\Rightarrow&\frac{\partial^{b}F_{ba_1\dotso a_{q-1}\dot a_1\dotso\dot a_{2p-q+1}}
+\partial^{\dot b}F_{a_1\dotso a_{q-1}\dot a_1\dotso\dot a_{2p-q+1}\dot b}}{(q-1)!(2p-q+1)!}=0,
\\
\vdots~~~~~~~~~~&&~~~~~~~~~~\vdots \nn \\
\frac{\de S}{\de A^{a_1\dotso a_{2q-k+1}\dot a_1\dotso\dot a_{2p-2q+k-1}}}=0
&\Rightarrow&\frac{\partial^{b}F_{ba_1\dotso a_{2q-k+1}\dot a_1\dotso\dot a_{2p-2q+k-1}}
+\partial^{\dot b}F_{a_1\dotso a_{2q-k+1}\dot a_1\dotso\dot a_{2p-2q+k-1}\dot b}}{(2q-k+1)!(2p-2q+k-1)!}=0, 
\nn \\
&& \label{oddsfd2}
\\
\notag \vdots~~~~~~~~~~&&~~~~~~~~~~\vdots\\
\frac{\de S}{\de A^{\dot a_1\dotso\dot a_{2p}}}=0
&\Rightarrow&\frac{\partial^{b}F_{b\dot a_1\dotso\dot a_{2p}}+\partial^{\dot b}F_{\dot a_1\dotso\dot a_{2p}\dot b}}{(2p)!}=0.
\eeqa

To solve \eqref{defodd1}, 
we substitute \eqref{sfdodd1} into \eqref{defodd1} and get
\beqa
\partial^{\dot a_{2p-q+2}}\cF_{a_1\dotso a_{q-1}\dot a_1\dotso\dot a_{2p-q+2}}=0. 
\eeqa
Its solution is 
\beqa
\cF_{a_1\dotso a_{q-1}\dot a_1\dotso\dot a_{2p-q+2}}=\frac{(-)^{q^2}
\epsilon_{a_1\dotso a_{2q+1}}\epsilon_{\dot a_1\dotso \dot a_{4p-2q+1}}\partial^{\dot a_{4p-2q+1}}
\Phi^{a_{q}\dotso a_{2q+1}\dot a_{2p-q+3}\dotso \dot a_{4p-2q}}}{(q+2)!(2p-q-2)!}.
\eeqa
Using a gauge transformation (\eqref{gtfodd} with $k=q-1$), 
\beq
A^{a_1\cdots a_{q+2}\dot{a}_1\cdots \dot{a}_{2p-q-2}} \rightarrow 
(A + \Phi)^{a_1\cdots a_{q+2}\dot{a}_1\cdots \dot{a}_{2p-q-2}},
\eeq
we obtain another self-duality condition
\beqa
\label{sfdodd2}\cF_{a_1\dotso a_{q-1}\dot a_1\dotso\dot a_{2p-q+2}}=0.
\eeqa

Now we can iterate this procedure to solve the equations of motion of 
$A^{a_1\dotso a_{q-2}\dot a_1\dotso\dot a_{2p-q+2}}$. 
In general, for a given value of $k$ from $q+2$ to $2q+1$, 
we solve an equation of motion by introducing arbitrary functions $\Phi$ as
\beqa
\notag \cF_{a_1\dotso a_{2q-k+1}\dot a_1\dotso\dot a_{2p-2q+k}}&=&\frac{(-)^{k^2}
\epsilon_{a_1\dotso a_{2q+1}}\epsilon_{\dot a_1\dotso \dot a_{4p-2q+1}}
\partial^{\dot a_{4p-2q+1}}\Phi^{a_{2q-k+2}\dotso a_{2q+1}\dot a_{2p-2q+k+1}\dotso \dot a_{4p-2q}}}{k!(2p-k)!}.
\label{oddD'Phi}
\eeqa
Then we obtain other self-duality conditions by using the gauge symmetry \eqref{gtfodd} 
from $k=q-1$ down to $k = 0$ iteratively. 

Finally, we have
\beq
\cF_{a_1\dotso a_{2q-k+1}\dot a_1\dotso\dot a_{2p-2q+k}}=0 
\qquad (k=q+1,\dotso, 2q+1).
\eeq 
These constitute all independent self-duality conditions for the gauge field.
That is, the Hodge dual of these conditions imply the other half of the self-duality conditions
\beq
\cF_{a_1\dotso a_{2q-k+1}\dot a_1\dotso\dot a_{2p-2q+k}}=0 
\qquad (k=0,\dotso, q).
\eeq

Similar to the case in section \ref{D'=D''},
the self-dual gauge fields $A_{SD}$ satisfying all self-duality conditions 
are related to the gauge potential $A$ in our theory only after 
field redefinitions (gauge transformations)
\beqa
A_{SD}^{a_1\dotso a_k\dot a_1\dotso\dot a_{2p-k}}&=&
A^{a_1\dotso a_k\dot a_1\dotso\dot a_{2p-k}}
\qquad (k=0,\dotso, q), \label{a11} \\
A_{SD}^{a_1\dotso a_k\dot a_1\dotso\dot a_{2p-k}} &=& 
(A+\Phi)^{a_1\dotso a_k\dot a_1\dotso\dot a_{2p-k}}
\qquad (k=q+1,\dotso, 2q+1), \label{a12}
\eeqa
where the fields $\Phi$ are introduced when solving the equations of motion
(see (\ref{oddD'Phi})).

\section{$D'' > D'$ $=$~even}
\label{evenDp}

The story for even $D'$ proceeds essentially in the same way as the odd $D'$ case, 
except that here we need to assume ${\cal M}_2^{D''}$ to be Euclidean 
(and so ${\cal M}_1^{D'}$ is Lorentzian).
Let $D' = 2r+2 < D''$ for some $r \in \{0,1\dotso,p-1\}$.
The action for this case is (\ref{S2}) 
\beqa
S=-\frac{1}{2}\int d^D x\sum_{j=r+1}^{2r+2}
\frac{\tilde F_{a_1\dotso a_j\dot a_1\dotso\dot a_{2p+1-j}}
\cF^{a_1\dotso a_j\dot a_1\dotso\dot a_{2p-j+1}}}{j!(2p-j+1)!2^{\de_{j,r+1}}},
\eeqa
and the gauge transformations are
\beqa
\label{gtfeven}\begin{array}{rll}
\de A^{a_1\dotso a_{2r-k+2}\dot a_1\dotso\dot a_{2p-2r+k-2}}=
&\Phi^{a_1\dotso a_{2r-k+2}\dot a_1\dotso\dot a_{2p-2r+k-2}}(x^a, x^{\dot a}),&~k=0,1,\dotso,r,\\
\de A^{a_1\dotso a_{2r-k+2}\dot a_1\dotso\dot a_{2p-2r+k-2}}=
&f^{a_1\dotso a_{2r-k+2}\dot a_1\dotso\dot a_{2p-2r+k-2}}(x^a),&~k=r+1,\\
\de A^{a_1\dotso a_{2r-k+2}\dot a_1\dotso\dot a_{2p-2r+k-2}}=&0,
&~k=r+2,\dotso,2r+2, 
\end{array}
\eeqa
in addition to the ordinary gauge transformations.
It is easy to check that $S$ is invariant under these transformations.

The nontrivial equations of motion of the gauge fields are 
\beqa
\label{VRAodd}
\frac{\de S}{\de A^{a_1\dotso a_{2r-k+2}\dot a_1\dotso\dot a_{2p-2r+k-2}}} = 0 
\qquad (k=r+1,\dotso,2r+2).
\eeqa
For $k=r+1$, 
the equation of motion is
\beq
\partial^{\dot a_{2p-r}}\cF_{a_1\dotso a_{r+1}\dot a_1\dotso\dot a_{2p-r-1}\dot a_{2p-r}} = 0, 
\label{k=r+1EOM}
\eeq
which is solved by
\beqa
\label{sfdsfdeven}
\cF_{a_1\dotso a_{r+1}\dot a_1\dotso\dot a_{2p-r-1}\dot a_{2p-r}}
=\frac{\epsilon_{a_1\dotso a_{2r+2}}\epsilon_{\dot a_1\dotso \dot a_{4p-2r}}
\partial^{\dot a_{2p-r+1}}\Phi^{a_{r+2}\dotso a_{2r+2}\dot a_{2p-r+2}\dotso \dot a_{4p-2r}}}{(r+1)!(2p-r)!}.
\eeqa
Taking the Hodge dual of both sides of this equation, we find
\beqa
\cF_{a_1\dotso a_{r+1}\dot a_1\dotso\dot a_{2p-r-1}\dot a_{2p-r}}=
\partial_{[\dot a_1}\Phi_{|a_1\dotso a_{r+1}|\dot a_2\dotso \dot a_{2p-r}]}. 
\eeqa  
Then the equation of motion (\ref{k=r+1EOM}) 
implies 
\beqa
\label{PhiDD}\partial^{\dot a_{2p-r}}\partial_{[\dot a_1}\Phi_{|a_1\dotso a_{r+1}|\dot a_2\dotso \dot a_{2p-r}]}=0.
\eeqa 
Now we notice that 
the field $\Phi$ defined via (\ref{sfdsfdeven}) is 
only defined up to a total derivative term
\beq
\Phi_{a_1\dotso a_{r+1}\dot a_2\dotso \dot a_{2p-r}}\rightarrow
\Phi_{a_1\dotso a_{r+1}\dot a_2\dotso \dot a_{2p-r}}
+\partial_{[\dot a_2}\La_{|a_1\dotso a_{r+1}|\dot a_3\dotso \dot a_{2p-r}]},
\eeq
which is similar in spirit to a gauge symmetry. 
(It is the gauge symmetry of a gauge symmetry.) 
Thus we can choose the Lorentz gauge for $\Phi$
\beq
\partial^{\dot a_2} \Phi_{a_1\dotso a_{r+1}\dot a_2\dotso \dot a_{2p-r}}=0.
\eeq
After gauge fixing, eq.\eqref{PhiDD} becomes
\beqa
\label{LapPhi}
\dot \partial^2 \Phi_{a_1\dotso a_{r+1}\dot a_1\dotso\dot a_{2p-r-1}}=0,
\eeqa
where $\dot\partial^2\equiv \del^{\dot a}\del_{\dot a}$ is the Laplacian on ${\cal M}_2^{D''}$. 
Assuming that ${\cal M}_2^{D''}$ is of Euclidean signature,
the only solution to the Laplace equation (\ref{LapPhi}) is
\footnotemark[7]
\footnotetext[7]{
Here we assume that the boundary condition of $\Phi$ at infinity 
is such that the integration $\int d^D x\; \del_{\dot{a}}(\Phi \del^{\dot a}\Phi)$ vanishes, 
so that,
for the inner product of two functions $f$ and $g$ defined by 
$\langle f| g\rangle \equiv \int d^D x\; f^* g$,
\beq
0 = -\langle \Phi | \dot{\del}^2 \Phi \rangle 
= \langle \del_{\dot a} \Phi | \del_{\dot a} \Phi \rangle \geq 0,
\eeq
which implies that $\del_{\dot a} \Phi = 0$.
}
\beqa
\Phi_{a_1\dotso a_{r+1}\dot a_1\dotso\dot a_{2p-r-1}}=
f_{a_1\dotso a_{r+1}\dot a_1\dotso\dot a_{2p-r-1}}(x^a) 
\eeqa
for some functions $f$ independent of $x^{\dot a}$. 
Hence (\ref{sfdsfdeven}) becomes a self-duality condition
\beqa
\label{sfdeq1}\cF_{a_1\dotso a_{r+1}\dot a_1\dotso\dot a_{2p-r-1}\dot a_{2p-r}}=0.
\eeqa

Next we consider the case $k=r+2$. 
The equation of motion (\ref{VRAodd}) for $k=r+2$ is
\beq
\label{sfdeven2}
\partial^{\dot b}
(\cF+F)_{a_1\dotso a_{r}\dot a_1\dotso\dot a_{2p-r}\dot b}+\partial^b 
F_{ba_1\dotso a_{r}\dot a_1\dotso\dot a_{2p-r}}=0.
\eeq
Using \eqref{sfdeq1}, we can rewrite the 2nd term in \eqref{sfdeven2}
as $\del^b \tilde{F}_{b\cdots}$, which equals $- \del^{\dot b} \tilde{F}_{\dot b\cdots}$ 
due to Bianchi identity.
As a result \eqref{sfdeven2} is equivalent to
\beqa
\partial^{\dot a_{2p-r+1}}\cF_{a_1\dotso a_{r}\dot a_1\dotso\dot a_{2p-r}\dot a_{2p-r+1}}=0,
\eeqa
and the solution is 
\beqa
\cF_{a_1\dotso a_{r}\dot a_1\dotso\dot a_{2p-r}\dot a_{2p-r+1}}=
\frac{\epsilon_{a_1\dotso a_{2r+2}}\epsilon_{\dot a_1\dotso \dot a_{4p-2r}}
\partial^{\dot a_{4p-2r}}\Phi^{a_{r+1}\dotso a_{2r+2}\dot a_{2p-r+2}\dotso \dot a_{4p-2r-1}}}{(r+2)!(2p-r-2)!}.
\eeqa
One can gauge away $\Phi$ on the right-hand side
by using the gauge degree of freedom in \eqref{gtfeven} with $k=r$, 
and arrive at the self-duality condition
\beqa
\label{sfdeq2}\cF_{a_1\dotso a_{r}\dot a_1\dotso\dot a_{2p-r+1}}=0.
\eeqa
The variation of the action with respect to the remaining gauge fields are
\beqa
\label{sfdeven0}
\frac{\de S}{\de A^{a_1\dotso a_{r-1}\dot a_1\dotso\dot a_{2p-r+1}}}=0
&\Rightarrow&\frac{\partial^{b}F_{ba_1\dotso a_{r-1}\dot a_1\dotso\dot a_{2p-r+1}}
+\partial^{\dot b}F_{a_1\dotso a_{r-1}\dot a_1\dotso\dot a_{2p-r+1}\dot b}}{(r-1)!(2p-r+1)!}=0,\\
\notag \vdots~~~~~~~~~~&&~~~~~~~~~~\vdots\\
\frac{\de S}{\de A^{a_1\dotso a_{2r-k+2}\dot a_1\dotso\dot a_{2p-2r+k-2}}}=0
\label{sfdeven3}
\notag&\Rightarrow&\frac{\partial^{b}F_{ba_1\dotso a_{2r-k+2}\dot a_1\dotso\dot a_{2p-2r+k-2}}
+\partial^{\dot b}F_{a_1\dotso a_{2r-k+2}\dot a_1\dotso\dot a_{2p-2r+k-2}\dot b}}{(2r-k+2)!(2p-2r+k-2)!}=0,\\
&&\\
\notag \vdots~~~~~~~~~~&&~~~~~~~~~~\vdots\\
\frac{\de S}{\de A^{\dot a_1\dotso\dot a_{2p}}}=0
&\Rightarrow&\frac{\partial^{b}F_{b\dot a_1\dotso\dot a_{2p}}
+\partial^{\dot b}F_{\dot a_1\dotso\dot a_{2p}\dot b}}{(2p)!}=0.
\eeqa
As we have done many times already, 
after substituting \eqref{sfdeq2} into \eqref{sfdeven0}, 
solving the differential equation, 
and then using the gauge symmetry \eqref{gtfeven} with $k=r-1$, 
we obtain another self-duality condition. 
Repeating the same manipulation for higher and higher values of $k$, 
one derives all the remaining self-duality conditions
\beqa
{\cal F}_{a_1\dotso a_{2r-k+2}\dot a_1\dotso\dot a_{2p-2r+k-1}}=0
\qquad (k=r+3,\dotso,2r+2).
\eeqa

Similar to the previous two sections, 
one can check that there is a 1-1 correspondence between 
the gauge potential $A$ in our theory and self-dual gauge configurations $A_{SD}$
\beqa
A_{SD}^{a_1\dotso a_k\dot a_1\dotso\dot a_{2p-k}}&=&
A^{a_1\dotso a_k\dot a_1\dotso\dot a_{2p-k}}
\qquad (k=0,\dotso, r+1), \label{a21} \\
A_{SD}^{a_1\dotso a_k\dot a_1\dotso\dot a_{2p-k}} &=& 
(A+\Phi)^{a_1\dotso a_k\dot a_1\dotso\dot a_{2p-k}}
\qquad (k=r+2,\dotso, 2r+2), \label{a22}
\eeqa
where $\Phi$ are the fields arising when we solve the equations of motion of $A$.

\section{Hidden Lorentz Symmetry}
\label{LorentzSymmetry}

The action $S_{D'+D''}$ (\ref{S1}) is manifestly invariant under 
the $SO(D'-1,1)\times SO(D'')$, or $SO(D') \times SO(D''-1,1)$ subgroup 
of the full Lorentz group $SO(D-1,1)$ of the spacetime 
${\cal M}^D = {\cal M}_1^{D'} \times {\cal M}_2^{D''}$, 
depending on the signature of ${\cal M}_1^{D'}$ and ${\cal M}_2^{D''}$. 
The Lorentz symmetry transformations which mix $x^a$ with $x^{\dot a}$ 
are no longer manifest.
Nevertheless, the action $S$ is actually fully Lorentz invariant 
under a modified Lorentz transformation rule for those transformations 
which mix $x^a$ with $x^{\dot a}$.
This is the main topic of this section. 

For simplicity we impose the gauge fixing condition
\beq
\label{gaugefix}
A_{a_1\dotso a_k\dot a_1\dotso\dot a_{2p-k}}=0
\qquad (k=\lfloor D'/2 \rfloor +1, \cdots, D'-\al), 
\eeq 
where $\lfloor s \rfloor$ is the largest integer not greater than $s$, 
$\al$ is given in (\ref{aldef}) so that $D' - \al = \mbox{min}(D', 2p)$.
Thus we only have to consider the transformations of the remaining fields
\beq
A_{a_1\cdots a_k\dot{a}_1\cdots \dot{a}_{2p-k}} \qquad 
(k = 0, 1, \cdots, \lfloor D'/2 \rfloor).
\eeq
Our claim is that 
\beq
\label{acLS}
\de S= \sum_{k=0}^{\lfloor D'/2 \rfloor} \frac{\de S}
{\de A_{a_1\dotso a_k\dot a_1\dotso\dot a_{2p-k}}}
\de A_{a_1\dotso a_k\dot a_1\dotso\dot a_{2p-k}}
\eeq
vanishes 
for the modified Lorentz transformation rule 
\beq
\label{mLor}
\de A_{a_1\dotso a_k\dot a_1\dotso\dot a_{2p-k}}
\equiv \de_1 A_{a_1\dotso a_k\dot a_1\dotso\dot a_{2p-k}} + 
\de_2 A_{a_1\dotso a_k\dot a_1\dotso\dot a_{2p-k}}, 
\eeq
where
\beqa
\de_1 A_{a_1\dotso a_k\dot a_1\dotso\dot a_{2p-k}}
&=&
k\la^{\dot b}_{[a_k}
A_{a_1\dotso a_{k-1}]\dot b\dot a_1\dotso\dot a_{2p-k}} 
-(2p-k)\la^b_{[\dot a_1}
A_{|a_1\dotso a_k b|\dot a_2\dotso\dot a_{2p-k-1}]} \nn \\
&&+\la^{b\dot b}(x_b\partial_{\dot b}-x_{\dot b}\partial_b)
A_{a_1\dotso a_k\dot a_1\dotso\dot a_{2p-k}}, \\
\notag\de_2 A_{a_1\dotso a_k\dot a_1\dotso\dot a_{2p-k}}&=&
\de_{\lfloor D'/2 \rfloor,k}\la^b_{\dot b}x^{\dot b}
\cF_{ba_1\dotso a_k\dot a_1\dotso\dot a_{2p-k}}\\
&&+\frac{1}{2}\de_{D'=even'}
\de_{\lfloor D'/2 \rfloor-1,k}\la^b_{\dot b}x^{\dot b}
\cF_{ba_1\dotso a_k\dot a_1\dotso\dot a_{2p-k}}.
\eeqa
In the expressions above, 
$k=0, 1,\cdots, \lfloor D'/2 \rfloor$,
and $\lam_{a\dot a}$ is the Lorentz transformation parameter.

Because of the gauge-fixing (\ref{gaugefix}), 
some of the field strength components vanish, 
and thus their dual field strengths also vanish
\beq
\label{ZVF}
F_{a_1\dotso a_{D^\prime-j}\dot a_1\dotso\dot a_{D/2-D^\prime+j}}
=\tilde F_{a_1\dotso a_{j}\dot a_1\dotso\dot a_{D/2-j}}=0
\eeq
for $j=0, 1, \cdots, \lceil D'/2 \rceil -2$.
Using the expressions of $\de S/\de A$ we computed earlier 
when we derived the equations of motion, 
we check after lengthy calculations that 
$\de_1 S + \de_2 S = 0$ for all decompositions $D = D' + D''$. 

More specifically, for $D' = D''$, 
\beq
\de_1 S = - \de_2 S = 
\frac{1}{2p!p!}\int d^D x \la^{b}_{\dot b}
\cF^{a_1\dotso a_p\dot a_1\dotso \dot a_p\dot b}\cF_{ba_1\dotso a_p\dot a_1\dotso \dot a_p};
\eeq
for $D' = 2q+1$, 
\beq
\de_1 S = - \de_2 S =
\frac{1}{2q!(2p-q)!}\int d^D x\la_{\dot b}^b\cF^{a_1\dotso a_{q}\dot a_1\dotso \dot a_{2p-q}\dot b}
\cF_{ba_1\dotso a_{q}\dot a_1\dotso \dot a_{2p-q}};
\eeq
and for $D' = 2r+2$,
\beq
\de_1 S = -\de_2 S =
\frac{1}{2r!(2p-r)!}\int d^D x\la^b_{\dot b}\cF^{a_1\dotso a_{r}\dot a_1\dotso\dot a_{2p-r}\dot b} 
\cF_{ba_1\dotso a_{r}\dot a_1\dotso\dot a_{2p-r}}.
\eeq

\section{Interaction with Charged Branes}

A $2p$-form gauge potential naturally couples to a $(2p-1)$-brane 
with a $2p$-dimensional worldvolume as the electric charge. 
The interaction between the $(2p-1)$-brane and the field 
should preserve all the gauge symmetry. 
In particular, the new gauge symmetry (\ref{gtf}) are needed 
in order to have first order differential equations as equations of motion.

We find the interaction term of the action to be given by
\beqa
\label{inter}
S_{D'+D''}^{int}=
-\int d^D x\sum_{j=\lceil \frac{D^{\prime}}{2}\rceil}^{D^{\prime}}
\frac{F_{a_1\dotso a_{D^{\prime}-j}\dot a_1\dotso\dot a_{D/2-D^{\prime}+j}}
\mathcal Q^{a_1\dotso a_{D^{\prime}-j}\dot a_1\dotso\dot a_{D/2-D^\prime+j}}}
{(D^{\prime}-j)!(\frac{D^{\prime\prime}-D^\prime}{2}+j)!2^{\de_{j,D'/2}}}, 
\eeqa
where 
\beqa
\mathcal Q=\si-\tilde \si
\eeqa
is a (2p$+$1)-form and $\si$ represents the source. 
Roughly speaking, the source term ($\si$ and its Hodge dual $\tilde{\si}$) should 
be determined by a Dirac delta function which specifies the location of the $(2p-1)$-brane. 
We will give explicit examples below.

With the interaction term (\ref{inter}) added, 
the modified equations of motion can again be reduced to 
first order differential equation using the additional gauge symmetries eq.\eqref{gtf} as
\beqa
\label{eomsoc3}(\cF+\mathcal Q)_{a_1\cdots a_k\dot a_1\cdots \dot a_{2p-k}\dot b}=0, 
\eeqa
saying that $F$ is no longer self-dual but $F + \sigma$ is.
At the same time the equations of motion imply
 \beqa
\label{eomsoc1} 
\partial^{\dot b}(\cF+\mathcal Q)_{a_1\cdots a_k\dot a_1\cdots \dot a_{2p-k}\dot b}=0,
\eeqa
By means of the Bianchi identity, we can rewrite eq.\eqref{eomsoc1} in following form 
which was obtained earlier in \cite{BelovMoore}:
\beqa
\label{eomsoc2}
\partial^{\dot b}(F+\si)_{a_1\cdots a_k\dot a_1\cdots \dot a_{2p-k}\dot b}
+\partial^{b}(\tilde F+\tilde \si)_{ba_1\cdots a_k\dot a_1\cdots \dot a_{2p-k}}
=\partial^\mu \tilde \si_{\mu a_1\cdots a_k\dot a_1\cdots \dot a_{2p-k}}.
\eeqa

Although it appears that the effect of the source term is merely 
to replace $F$ by $F + \si$ in both equations of motion \eqref{eomsoc3} and \eqref{eomsoc1},  
one cannot redefine $F$ to absorb $\si$. 
The reason is that, 
although $F$ and $\si$ are both $(2p+1)$-forms, 
$F$ is a closed differential form while $\si$ is not. 
This also explains why there is an additional term on the right-hand side of (\ref{eomsoc2}).

To be more explicit, let us give some examples for each type of spacetime decomposition:
\begin{enumerate}
\item $D'=D''$
\beqa
\label{soc1}
\partial^{\mu} \tilde \si_{x_1\cdots x_k\dot x_1\cdots \dot x_{2p-k}\mu}
&=&\prod^{2p+1}_{\ell=k+1}\de(x_{\ell})\prod^{2p+1}_{\ell=2p-k+1}\de(\dot x_{\ell})
\qquad
(0 \leq k \leq p).
\eeqa
\item $D' =$ odd
\beqa
\label{soc2}
\partial^{\mu} \tilde \si_{x_1\cdots x_k\dot x_1\cdots \dot x_{2p-k}\mu}
&=&\prod^{2q+1}_{\ell=k+1}\de(x_{\ell})\prod^{4p-2q+1}_{\ell=2p-k+1}\de(\dot x_{\ell})
\qquad
(0 \leq k \leq q).
\eeqa
\item $D' =$ even
\beqa
\label{soc3}
\partial^{\mu} \tilde \si_{x_1\cdots x_k\dot x_1\cdots \dot x_{2p-k}\mu}
&=&\prod^{2r+2}_{\ell=k+1}\de(x_{\ell})\prod^{4p-2r}_{\ell=2p-k+1}\de(\dot x_{\ell}) 
\qquad
(1\leq k \leq r+1).
\eeqa
\end{enumerate}
Here the value of $k$ is a fixed number used to 
specify how the $2p$-dimensional worldvolume of the source brane 
is divided into ${\cal M}_1^{D'}\times {\cal M}_2^{D''}$.
We use the divergence of $\tilde{\si}$ to define $\si$, 
with all other components of the divergence of $\tilde{\si}$ vanishing. 
These divergence relations do not determine $\si$ uniquely 
but different solutions of $\si$ are equivalent.\footnote{
A more familiar expression of the interaction term is of the form $A\cdot j$ 
for some current $j$.
In (\ref{inter}) the interaction term is of the form $- F\cdot Q$, 
and since $F = dA$, 
it is equivalent to $A\cdot dQ$ up to a total derivative. 
Thus $dQ$ is the current $j$ coupled to $A$.
}
The $x_\ell$'s and $\dot x_\ell$'s appearing in the delta functions 
(there are $2p+2$ of them)
are the coordinates transverse to the brane. 

In order for the brane to be time-like, 
one should choose the non-vanishing components of $\partial \cdot \tilde\si$ to have a time-like index. 
This imposes a constraint on the choice of $\partial \cdot \tilde \si$ for the $D'=$ even case because 
the coordinate space $\{x^a\}$ of ${\cal M}_1^{D'}$ is of Lorentzian signature.

For the source given above, 
the equation of motion \eqref{eomsoc2} becomes
\beqa
\label{eomsoc4}&&
\partial^{\dot b}(F+\si)_{x_1\cdots x_k\dot x_1\cdots \dot x_{2p-k}\dot b}
+\partial^{b}(\tilde F+\tilde \si)_{bx_1\cdots x_k\dot x_1\cdots \dot x_{2p-k}}\\
\notag &=&\left\{\begin{array}{lll}
\dps\prod^{2p+1}_{\ell=k+1}\de(x_{\ell})\prod^{2p+1}_{\ell=2p-k+1}\de(\dot x_{\ell}),
& (D'=D'') ,\\
\dps\prod^{2q+1}_{\ell=k+1}\de(x_{\ell})\prod^{4p-2q+1}_{\ell=2p-k+1}\de(\dot x_{\ell}),
& (D'= \mbox{odd}) ,\\
\dps\prod^{2r+2}_{\ell=k+1}\de(x_{\ell})\prod^{4p-2r}_{\ell=2p-k+1}\de(\dot x_{\ell}),
& (D'= \mbox{even}) .
\end{array}\right.
\eeqa

\section{Examples}
\label{Examples}

\subsection{\large $D=2$, $(D', D'') = (1, 1)$}

The simplest example of self-dual field theory 
is the two-dimensional theory of chiral bosons. 
The only possible decomposition of the 2D Minkowski space is 
\beq
\mathbb{R}^{1+1} = \mathbb{R} \times \mathbb{R}.
\eeq
We use $x^1$ and $x^{\dot 1}$ to denote the coordinates on each factor of $\mathbb{R}$. 
Either $x^1$ or $x^{\dot 1}$ can be time-like, and the other one space-like.

The gauge potential has no index and has only one component 
which will be denoted as $\phi$. 
The action is
\beq
S_{1+1}=-\frac{1}{2}\int d^2 x F^{\dot 1}\cF_{\dot 1} ,
\eeq
where $F_{\dot 1}=\partial_{\dot 1}\phi$ and $\tilde F^{\dot 1}=\partial_{1}\phi$.
It is the same as the action given in \cite{FJ}.
Note that this action is invariant under the gauge transformation
(the second line in (\ref{gtf}))
\beq
\label{gaugephi}
\phi \rightarrow \phi + f(x^1)
\eeq
for an arbitrary function $f(x^1)$ that is independent of $x^{\dot{1}}$.
This is a larger gauge symmetry than the usual gauge symmetry 
for a 0-form gauge potential,
which is restricted to constant $f$.

The action is not manifestly Lorentz invariant. 
Define the modified Lorentz transformation law
\beq
\delta \phi = \lambda^{a\dot{a}} (x_a \del_{\dot{a}} - x_{\dot{a}} \del_a) \phi
+ \lambda^a_{\dot{a}} x^{\dot{a}} {\cal F}_a 
\equiv \delta_1 \phi + \delta_2 \phi, 
\eeq
where 
\beq
\delta_1 \phi = \lambda^{a\dot{a}} (x_a \del_{\dot{a}} - x_{\dot{a}} \del_a) \phi, 
\qquad
\delta_2 \phi = \lambda^a_{\dot{a}} x^{\dot{a}} {\cal F}_a. 
\eeq
When ${\cal F} = 0$, $\delta_2 \phi = 0$ and 
$\delta_1 \phi$ is the usual Lorentz transformation law.
Then one can check that 
\beq
\delta_1 S = - \delta_2 S = 
\frac{1}{2} \int d^2 x \; \lambda^{\dot{a}}_a {\cal F}_{\dot{a}} {\cal F}^a 
\eeq
and so the action $S$ is invariant under this modified Lorentz transformation law.

The equation of motion is 
\beq
\frac{\de S_{1+1}}{\de \phi}=0~\Rightarrow~\partial^{\dot 1}\cF_{\dot 1}=0,
\eeq
whose solution is
\beq
\cF_{\dot 1}=g(x^1),
\eeq
where $g(x^1)$ is an arbitrary function depending only on $x^1$.
The function $g(x^1)$ can be absorbed by the gauge transformation (\ref{gaugephi})
with $f(x^1) = - \int^{x^1} dy g(y)$,
and then we have the self-duality condition
\beq
\cF_{\dot 1}=0.
\eeq

\subsection{$D=6$, $(D', D'') = (1, 5)$}

The decomposition $(D', D'') = (1, 5)$ was used previously 
for the formulation of M5-brane \cite{oldM5}
and type \IIB supergravity \cite{IIB9+1}.
The gauge potential has the following components
\beq
A_{1\dot{a}}, \quad A_{\dot{a}\dot{b}} 
\qquad (\dot{a}, \dot{b} = \dot{1}, \cdots, \dot{5}).
\eeq
The action is 
\beq
S_{1+5} = -\frac{1}{4}\int d^6 x \tilde F_{1\dot a\dot b}\cF^{1\dot a\dot b}.
\eeq
The equation of motion obtained by varying $S_{1+5}$ with respect to $A^{1\dot{a}}$ is
\beq
\frac{\de S_{1+5}}{\de A^{1\dot{a}}} =0~\Rightarrow~ 
\partial^{\dot b}\tilde F_{1\dot a\dot b} \equiv 0, 
\eeq
which is identically zero. 
This means that $S_{1+5}$ depends on $A^{1\dot{a}}$ only through 
total derivative terms. 
Thus the theory has the gauge symmetry 
\beq
\label{gaugeA1adot}
A^{1\dot{a}} \rightarrow \Phi^{1\dot{a}}
\eeq
for arbitrary functions $\Phi^{1\dot{a}}$.

The equation of motion for $A_{\dot{a}\dot{b}}$ is
\beq
\label{eom1+5}\dps
\frac{\de S_{1+5}}{\de A^{\dot{a}\dot{b}}} =0~\Rightarrow~ 
\partial^{\dot c}\cF_{\dot a\dot b\dot c}=0,
\eeq
whose solution is 
\beq
\cF_{\dot a\dot b\dot c}=\epsilon_{1\dot a\dot b\dot c\dot d\dot e}\partial^{\dot d}\Phi^{1\dot e},
\eeq
where $\Phi^{1\dot e}$ are arbitrary functions.
Using the gauge transformation of $A_{1\dot a}$ (\ref{gaugeA1adot}), 
we can make a gauge transformation to absorb $\Phi^{1\dot e}$ 
and arrive at the self-duality condition
\beq
\label{cFabcdots}
\cF_{\dot a\dot b\dot c}=0,
\eeq 
which is the only one self-duality condition in the special case.

An equivalent formulation of this theory is to start with a theory without $A^{1\dot{a}}$. 
The only fields of this alternative formulation are $A^{\dot{a}\dot{b}}$. 
The action is taken to be $S_{1+5}$ with all $A^{1\dot{a}}$ set to zero, 
so that $F_{1\dot{d}\dot{e}}$, which appears in ${\cal F}_{\dot{a}\dot{b}\dot{c}}$,
is now replaced by 
\beq
f_{1\dot{d}\dot{e}} = \del_1 A_{\dot{d}\dot{e}}.
\eeq
The equation of motion of $A^{\dot{a}\dot{b}}$ is still of the form (\ref{eom1+5}), 
but with $\cF_{\dot a\dot b\dot c}$ replaced by 
\beq
F_{\dot a\dot b\dot c} - \frac{1}{2} \epsilon_{\dot a\dot b\dot c\dot d\dot e} f^{1\dot d\dot e}
= F_{\dot a\dot b\dot c} + \frac{1}{2} \epsilon_{\dot a\dot b\dot c\dot d\dot e}\partial^1 A^{\dot d\dot e}
= \epsilon_{1\dot a\dot b\dot c\dot d\dot e}\partial^{\dot d}\Phi^{1\dot e}.
\eeq
This can be rewritten as the self-duality condition (\ref{cFabcdots})
if we define 
\beq
F_{1\dot{d}\dot{e}} \equiv \del_1 A_{\dot{d}\dot{e}} - \del_{\dot{d}} \Phi_{1\dot{e}} + \del_{\dot{e}} \Phi_{1\dot{d}}. 
\eeq
Hence the functions $\Phi^{1\dot{a}}$ appearing in the solution of the equations of motion 
should be interpreted as the components $A^{1\dot{a}}$ of a self-dual gauge potential 
together with $A^{\dot{a}\dot{b}}$.

\subsection{$D=6$, $(D', D'') = (2, 4)$}

Instead of decomposing the 6D Minkowski space $\mathbb{R}^{1+5}$ 
as $\mathbb{R}^1\times \mathbb{R}^5$ or $\mathbb{R}^1\times \mathbb{R}^{1+4}$ 
like we did in the previous subsection, 
here we consider another decomposition
\beq
\mathbb{R}^{1+5} = \mathbb{R}^{1+1}\times \mathbb{R}^4.
\eeq
To our knowledge 
this decomposition was never discussed in the literature.
The 2-form gauge potential has the following components
\beq
A_{ab},\quad A_{a\dot{a}},\quad A_{\dot{a}\dot{b}} \qquad
(a, b = 1, 2; \;\; \dot{a}, \dot{b} = \dot{1}, \cdots, \dot{4}).
\eeq
The action is 
\beq
\label{S2+4}
S_{2+4}=-\frac{1}{8}\int d^6 x \big(2\tilde F_{ab\dot c}
\cF^{ab\dot c}+\tilde F_{a\dot b\dot c}\cF^{a\dot b\dot c}\big).
\eeq 
The variation of the action with respect to $A_{ab}$,
\beq
\frac{\de S_{2+4}}{\de A^{ab}}=0~\Rightarrow~\frac{1}{4}\partial^{\dot a}\tilde F_{ab\dot a}= 0,
\eeq
vanishes identically, 
and thus the gauge potential $A_{ab}$ can perform the gauge transformation
\beq
A_{ab} \rightarrow A_{ab} + \Om_{ab}.
\eeq 
The equations of motion of $A_{a\dot a}$ and $A_{\dot a\dot b}$ are,
\beqa
\label{eom2+41}\dps
\frac{\de S_{2+4}}{\de A^{a\dot a}}=0&\Rightarrow&\frac{1}{2}\partial^{\dot b}\cF_{a\dot a\dot b}=0, \\
\label{eom2+42}\dps
\frac{\de S_{2+4}}{\de A^{\dot a\dot b}}=0&\Rightarrow&\frac{1}{4}
\Big[\partial^{\dot c}(\cF+F)_{\dot a\dot b\dot c}+\partial^a F_{a\dot a\dot b}\Big]=0.
\eeqa
The solution to eq.\eqref{eom2+41} is 
\beqa
\label{soleom2+411}
\cF_{a\dot a\dot b}=\epsilon_{ab}\epsilon_{\dot a\dot b\dot c\dot d}\partial^{\dot c}\Phi^{b	\dot d}
\eeqa
for some arbitrary functions $\Phi^{b \dot d}$.
Taking the Hodge-dual of both sides of eq.\eqref{soleom2+411}, 
we have another solution to eq.\eqref{eom2+41}
\beqa
\cF_{a\dot a\dot b}=\partial_{\dot a}\Phi_{a\dot b}-\partial_{\dot b}\Phi_{a\dot a}.
\eeqa
Identifying these two solution of eq.\eqref{eom2+41} leads to  
\beqa
\label{phi3}
\partial_{\dot a}\Phi_{a\dot b}-\partial_{\dot b}\Phi_{a\dot a}=
\epsilon_{ab}\epsilon_{\dot a\dot b\dot c\dot d}\partial^{\dot c}\Phi^{b	\dot d},
\eeqa 
and then we act $\partial^{\dot a}$ on both sides of the equivalence relation,
\beqa
\label{phi4}
\partial^{\dot{b}}\partial_{\dot{b}}\Phi_{a\dot a}-\partial_{\dot a}\partial^{\dot b}\Phi_{a\dot b}=0.
\eeqa
There are two gauge degrees of freedom in the field $\Phi^{a\dot a}$ 
because the content of the equation (\ref{soleom2+411}) is unchanged 
under the transformation 
\beq
\Phi_{a\dot a}\rightarrow \Phi_{a\dot a}+\partial_{\dot a}\La_a.
\eeq
As a result we can choose the Lorentz gauge 
$\partial^{\dot a}\Phi_{a\dot a}=0$ 
and then \eqref{phi4} reduces to
\beqa
\dot \partial^2\Phi_{a\dot a}=0,
\label{Laplacephiaad}
\qquad 
\mbox{where} \qquad
\dot \partial^2\equiv \partial^{\dot a}\partial_{\dot a}.
\eeqa  
Imposing the boundary condition that $\Phi^{a\dot a}$ vanishes at infinities
of the 4D Euclidean space $\mathbb{R}^4$ 
such that (\ref{Laplacephiaad}) has the unique solution 
$\Phi_{a\dot a}=0$, 
we arrive at one of the self-duality conditions
\beq
\label{2+4sfd1}
\cF_{a\dot a\dot b}=0.
\eeq

Next, plugging \eqref{2+4sfd1} into \eqref{eom2+42} 
implies that
\beqa
\cF_{\dot a\dot b\dot c}=\frac{1}{2}\epsilon_{ab}\epsilon_{\dot a\dot b\dot c\dot d}\partial^{\dot d}\Om^{ab}
\eeqa
for some arbitrary functions $\Om^{ab}$.
After performing the gauge transformation $\de A_{ab}=\Om_{ab}$,
the equation above becomes
the other self-duality condition
\beq
\label{2+4sfd2}
\cF_{\dot a\dot b\dot c}=0.
\eeq
Eqs. (\ref{2+4sfd1}) and (\ref{2+4sfd2}) are sufficient to guarantee 
the self-dualty conditions for all components of the field strength. 

To summarize, 
solutions of $A$ to the equations of motion derived from the action $S_{2+4}$ (\ref{S2+4}) 
can be identified with a self-dual gauge potentials $A_{SD}$
\beq
A_{SD}^{ab} =(A+\Om)^{ab}, \qquad 
A_{SD}^{a\dot b} = A^{a\dot b}, \qquad 
A_{SD}^{\dot{a}\dot{b}} = A^{\dot{a}\dot{b}}. 
\eeq

\subsection{$D=6$, $(D', D'') = (3, 3)$}

Now we introduce the 3rd formulation for 6D self-dual gauge theory
based on the decomposition of 6D Minkowski space
\beq
\mathbb{R}^{1+5} = \mathbb{R}^{1+2} \times \mathbb{R}^{3}.
\eeq
This formulation was first introduced in \cite{Ho:2008nn} 
as a linearized version of the M5-brane theory. 
It was later extended to the full interacting M5-brane theory in \cite{Ho:2008ve}, 
and the self-duality conditions in the nonlinear version 
of the pure gauge theory
were explicitly examined in \cite{Pasti}, 
and the full theory including interactions between gauge fields 
and matter fields is analyzed in \cite{Kazu}.

The gauge potential has the following components
\beq
A_{ab}, \quad A_{a\dot{a}}, \quad A_{\dot{a}\dot{b}} \qquad
(a, b = 1, 2, 3; \;\; \dot{a}, \dot{b} = \dot{1}, \dot{2}, \dot{3}),
\eeq
where the indices $a = (1, 2, 3)$ and ��$\dot a = (\dot 1,\dot 2,\dot 3)$, correspond, respectively, to the $SO(2,1)$ and
$SO(3)$ subgroup of the full $D = 6$ Lorentz group.
The action is 
\beq
S_{3+3}=-\frac{1}{12}\int d^6 x \big(\tilde F_{abc}\cF^{abc}+3\tilde F_{ab\dot a}\cF^{ab\dot a}\big).
\eeq
The terms involving $A_{ab}$ is a total derivative, 
and thus the equation of motion of $A_{ab}$ is trivial
\beq
\frac{\de S_{3+3}}{\de A^{ab}}=0~\Rightarrow~
\frac{1}{4}\big(\partial^{c}\tilde F_{abc}
+\partial^{\dot c}\tilde F_{ab\dot c}\big)= 0.
\eeq
This means that 
\beq
A_{ab} \rightarrow A_{ab} + \Phi_{ab}
\label{gaugeAab3+3}
\eeq
is a gauge symmetry.

The equations of motion of $A_{a\dot a}$ and $A_{\dot a\dot b}$ are
\beqa
\label{eom3+31}\dps\frac{\de S_{3+3}}{\de A^{a\dot a}}=0&\Rightarrow&\partial^{\dot b}\cF_{a\dot a\dot b}=0,\\
\label{eom3+32}\dps\frac{\de S_{3+3}}{\de A^{\dot a\dot b}}=0&\Rightarrow&\frac{1}{2}\big(\partial^a F_{a\dot a\dot b}+\partial^{\dot a} F_{\dot a\dot b\dot c}\big)=0.
\eeqa
The solution to (\ref{eom3+31}) is 
\beq
\cF_{a\dot a\dot b}=\frac{1}{2}\epsilon_{abc}\epsilon_{\dot a\dot b\dot c}\partial^{\dot c}\Phi^{bc}
\eeq
for some arbitrary functions $\Phi^{bc}$.
Through a suitable gauge transformation (\ref{gaugeAab3+3}), 
it can be reduced to
\beq
\label{sfd3+31}
\cF_{a\dot a\dot b}=0.
\eeq
Substituting \eqref{sfd3+31} into \eqref{eom3+32}, 
we solve the latter as
\beq
\cF_{\dot a\dot b\dot c}=\epsilon_{\dot a\dot b\dot c}f(x^a),
\eeq
where $f(x)$ is a function depending only on the coordinates $x^a$ 
but not on $x^{\dot a}$. 
Since $f(x)$ can always be written as $\partial^{a}f_a$ for some fields $f_a(x^b)$, 
it can be absorbed by a field redefinition $A_{ab}\rightarrow A_{ab}+\epsilon_{abc}f^c$. 
One can check that this does not spoil the self-duality condition \eqref{sfd3+31} 
we obtained earlier. 
Hence we finally have the other self-duality condition,
\beq
\cF_{\dot a\dot b\dot c}=0.
\eeq
Together with (\ref{sfd3+31}), 
they constitute the full self-duality conditions for the 6D self-dual gauge field $A$.

\section{Comparison with the holographic action}
\label{Compare}

The results presented above are closely related to the work of Belov and Moore \cite{BelovMoore}. 
\footnote{
We thank Yuji Tachikawa for pointing out this reference to us 
after the first version of this paper appeared on arXiv.
}
They defined an action, which is called the holographic action, for self-dual gauge fields, 
based on a decomposition of the space of $(2p+1)$-forms $V_{\mathbb{R}}$. 
In their notation, 
$V_{\mathbb{R}} = V_2 \oplus V_2^{\perp}$, 
where $V_2^{\perp} = \ast_E V_2$ is the orthogonal complement 
with respect to the Hodge metric. 
(The subscript $E$ stands for {\em Euclidean}, 
as the formulation starts with a Wick-rotated space
with Euclidean signature.)
Thus, in particular, a field strength $F$ can be decomposed as 
$F = F_2 + F_2^{\perp}$.
($F$ is denoted as $R$ in \cite{BelovMoore}.) 

The decomposition of spacetime ${\cal M}^D = {\cal M}_1^{D'} \times {\cal M}_2^{D''}$ 
we considered in this paper also leads to a decomposition of 
the space of differential forms. 
For $(2p+1)$-forms, 
we can choose 
\beq
V_2^{\perp} = \{ 
dx^{a_1}\cdots dx^{a_{k}} dx^{\dot{a}_1}\cdots dx^{\dot{a}_{2p-k+1}} | 
k = 0, 1, \cdots, \lfloor D'/2\rfloor \}.
\eeq
For a given field strength $F$, 
the part of $F_2^{\perp}$ depends only on 
the components 
\beq
\{ A_{a_1\cdots a_{k}\dot{a}_1\cdots \dot{a}_{2p-k}} | k = 0, 1, \cdots, \lfloor D'/2\rfloor \}
\eeq
of the gauge potential, 
and these are exactly the components of the gauge potential 
that cannot be gauged away in our formulation 
(see, e.g, (\ref{gtf})). 
The decomposing the space of the field strength $F$ 
as in the work of Belov and Moore is thus implied by 
our classification of the gauge potential components $A$. 

Another technical difference between these two approaches 
is that we used the equations of motion to prove the self-duality condition, 
while Belov and Moore used the equations of motion to prove 
that the $(2p+1)$-form
$F^+ \equiv F_2^{\perp} + \ast F_2^{\perp}$, 
which is by definition self-dual, is closed. 
Nevertheless, since the self-dual field strength is 
completely fixed by the part $F_2^{\perp}$ in $V_2^{\perp}$, 
the full field strength is the same independent of 
whether the rest of the components are obtained by definition 
as $\ast F_2^{\perp}$, 
or obtained by solving equations of motion
which lead to self-duality conditions. 

Apart from these technical details, 
our work can be viewed as an application of the holographic action \cite{BelovMoore}
with a special choice of the decomposition of $V_{\mathbb{R}}$. 
The work of Belov and Moore \cite{BelovMoore} 
is more general and has the advantage of being able to deal with 
topological issues more properly. 
They have also shown how to couple D-branes to the self-dual fields.
Our work is on the other hand focused on a special realization of their general formulation, 
giving explicit expressions of the action in terms of gauge potential components, 
and explicit expressions for Lorentz transformation laws.
This is achieved by realizing that the construction of a self-dual field theory 
is naturally connected with the decomposition of space-time,
when it is a product space.

\section{Conclusion}
\label{Conclusion}

In this work we constructed Lagrange formulations 
for self-dual gauge theories in $D = 4p+2$ dimensions ($p = 0, 1, 2, \cdots$)
for arbitrary decomposition $D = D' + D''$. 
In addition to the ordinary gauge symmetry 
\beq
A^{(2p)} \rightarrow A^{(2p)} + d\Lambda^{(2p-1)},
\eeq
the action (\ref{S1})-(\ref{S3}) is invariant under additional gauge transformations, 
which allow us to gauge away some of the components of $A$, 
leaving the following components
\beq
A_{a_1\dotso a_{k}\dot a_1\dotso\dot a_{2p-k}} 
\qquad (k=0,1, \cdots, \lfloor D'/2 \rfloor)
\label{remcomp}
\eeq
as dynamical variables satisfying 2nd order differential equations. 
In the process of solving the equations of motion for these remaining components of $A$, 
additional fields $\Phi$ are introduced when 2nd order differential equations 
are integrated to 1st order differential equations. 
Together with those components of $A$ in (\ref{remcomp}), 
the fields $\Phi$ are to be identified with the rest of the components 
of the self-dual configuration.
Despite the fact that the action is not manifestly Lorentz invariant, 
it enjoys the full Lorentz symmetry with a modified transformation law (\ref{mLor}).

Potential applications of our new formulations of self-dual gauge theories
include rewriting a self-dual gauge theory, 
say, the action of type \IIB supergravity for decompositions
other than $(D', D'') = (1, 9)$, 
which was already constructed in \cite{IIB9+1}
Other decompositions $(D', D'')$ may be more convenient 
when specific backgrounds are considered, 
e.g. when the spacetime is naturally viewed as a product space.
For example, 
the decomposition $(D', D'') = (5, 5)$ will be especially convenient 
when we try to construct new solutions of \IIB supergravity 
as deformations of the $AdS_5\times S^5$ background.

\subsection*{Acknowledgments}

The authors thank Chien-Ho Chen, Kazuyuki Furuuchi, 
Yu-Tin Huang, Sheng-Lan Ko, Sangmin Lee, Paolo Pasti,
Dmitri Sorokin, Yuji Tachikawa, 
Tomohisa Takimi, Daniel Thompson and Chi-Hsien Yeh. 
This work is supported in part by
the National Science Council,
and the National Center for Theoretical Sciences, Taiwan, R.O.C. 


\end{document}